# STELLAR POPULATIONS OF BULGES AND DISKS: NEW INSIGHTS FROM NEAR-IR COLOURS

R. F. PELETIER

*Kapteyn Astronomical Institute*
*Postbus 800, 9700 AV Groningen, Netherlands*
*and*
*Instituto de Astrofísica de Canarias*
*Via Lactea s/n, 38200 La Laguna, Tenerife, Spain*

AND

M. BALCELLS

*Kapteyn Astronomical Institute*
*Postbus 800, 9700 AV Groningen, Netherlands*

**Abstract.** Colours of spiral galaxies generally are affected by the underlying old stellar population, younger stars, and extinction by dust. Old and young populations generally can be disentangled using a combination of blue and near-infrared colours. Extinction effects are very hard to take into account, except for galaxies with special orientations. In this paper we give some results of one of the first studies of galactic bulges and disks in various optical and near-infrared bands, and its implications for the stellar populations of spirals.

## 1. Introduction

When one compares spiral galaxies with ellipticals, one can notice that much less is known about the stellar populations of the former than of the latter, When one then looks at a grand-design spiral, one immediately understands the reason for this. Spirals not only seem to have an old, underlying population, but many times they also contain young stars, as well as dust between the stars. For ellipticals one can in general get a good idea of the metallicity, just from an optical colour like $B - V$. For spirals with a combination of young and old stellar populations one colour is not sufficient to be able to separate them. Even a combination of several optical colours (e.g. $U - V$ and $B - V$) is generally not sufficient, since one cannot distinguish between a lower metallicity of the old population, and a larger fraction of young stars. To overcome this problem one has to go to the near-infrared. Tully *et al.* (1982) and especially Frogel (1985) has shown that the $U - V$ vs. $V - K$ diagram is a very useful tool for detecting young stellar populations in spirals. Since $V - K$ is a very red colour, it happens that young stars do not contribute as much in $V - K$ as in $U - V$, and that for a given $V - K$ the $U - V$ of an old population is very different from the $U - V$ of a combination of old and young stars. On the other hand Frogel noticed that there is a considerable scatter in the $U - V$ vs. $V - K$ diagram for spirals, much larger than for ellipticals, and that many spirals were redder than the reddest ellipticals.

These two effects very likely are due to extinction by dust. Images of edge-on galaxies, like NGC 891, show many magnitudes of extinction by dust in the central regions of the disk. Because of this, the integrated colours of this galaxy are very red (Wirth & Shaw 1983). In Balcells & Peletier (1994,hereinafter called Paper I) we have shown that the colours of the stellar populations themselves in these objects are not red, in general even bluer than those of ellipticals of the same size.

Since that paper only discusses data in the optical, the information presented about stellar populations in bulges was limited. For that reason we have obtained images in $J$ and $K$ for the sample, with a new two-dimensional detector with a resolution that is sufficient to spatially select regions that are not or almost not affected by extinction. The colours (optical and optical-infrared) of these regions have been analyzed in the way described by Frogel (1985), and in this paper we discuss the implications for stellar populations in spiral galaxies.

## 2. Sample and Observations

We have investigated a sample of early-type spiral galaxies, ranging from S0's to Sbc. These galaxies generally have bulges that can be separated rather easily from their surrounding disk. Furthermore, except for the latest type, the amount of extinction by dust in these galaxies is limited, so that the colours will still be able to contain information about the stellar populations. The galaxies have inclinations larger than 50°, and are the brightest galaxies in a certain part of the sky. These two properties enable us to also on the basis of their morphology, and not just on their surface brightness profiles, separate disk and bulge.

The sample that we observed were are galaxies of Table 1 of Paper I, except for 2, which had declinations larger than 60°, and so could not be observed at UKIRT. Optical data (presented in that paper) were obtained for all galaxies. All galaxies of the *dustfree* subsample were observed both in $J$ and $K$. The other galaxies were all observed in $K$, but not necessarily in $J$.

The optical data consist of $U, B, R$ and $I$ surface photometry, obtained in June, 1990, on the 2.5m INT telescope at La Palma. The data have been described in Paper I. The pixelsize of the data was 0.549", and the effective seeing on the images lies between 1.2" and 1.6". The images were taken under photometric conditions.

The near-infrared data were observed in June, 1994, at the 4m United Kingdom Infrared Telescope at Hawaii, using IRCAM3 (Puxley & Aspin 1994), an infrared camera equipped with a 256 × 256 InSb array. The observations will be described in detail in a subsequent paper (Peletier & Balcells, in preparation). The array has a pixelsize of 0.291", as measured on the images, so that the field is about 75" on the side. Cosmetically, the array is very clean, with less than 1% bad pixels. For every object we took images in 10 positions, which each consisted of several readouts, making up a total integration time of 100s per position. The object was moved around on the chip on 6 of these exposures, while 4 consisted of blank sky, about 10' from the galaxy. The data were flatfielded using median sky flat fields, and a final mosaic was made aligning the individual frames. The effective seeing on the final frames was between 0.8" and 1.0". Here also the frames were taken under photometric conditions, with maximum zero point errors of 0.1 mag in $J$ and $K$.

In Fig. 1 we show for a typical galaxy a composite $U - R - K$ map, and two colour maps. These show the bulge and disks, and the regions with the largest extinction and star formation. In Fig. 2 we show a cut along the minor axis in the $U - R$ vs. $R - K$ diagram. One side of the galaxy is well behaved, and $U - R$ is increasing for increasing $R - K$, but on the other side the combination of extinction, scattering and star formation makes this diagram very hard to interpret.

Figure 1. Greyscale image and two colour maps of IC 1029, showing the geometry and position of the dustlanes in a typical galaxy (a) and the lack of colour differences between bulges and disks (b and c).

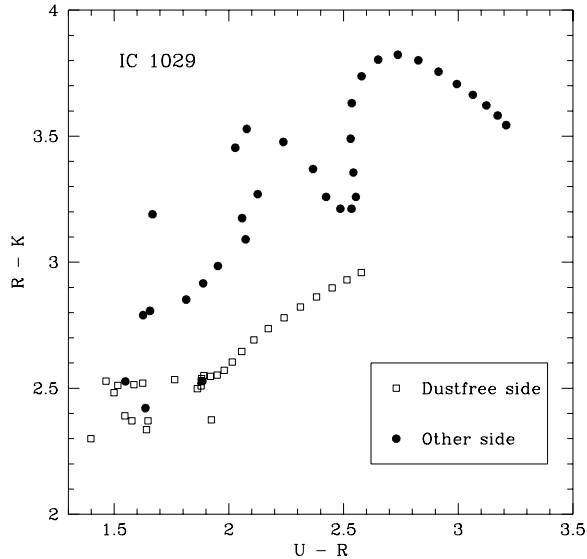

Figure 2. Color-color diagram of a minor axis cut of IC 1029. Both sides are indicated by different symbols.

## 3. Dust in spiral galaxies

Colours contain useful information about stellar populations only when they have been corrected for extinction. Since the effects of extinction can be severe, the errors introduced by correcting for it are often so large that it is better to choose regions of the galaxy that are not, or very little, affected by extinction. In paper I we describe the procedure we employ to find extinction-free regions of bulges. For disks the situation is more complicated. Since dust and stars here are mixed almost everywhere, one cannot find dustfree regions. From a colour map (e.g. Fig. 1) one can see that the extinction in these early-type spirals is almost all concentrated on one side of the galaxy. This is the case for all galaxies that are not seen close to edge-on. One minimizes the extinction if one measures the disk colours on the other side. Since we are interested in radial and not vertical disk profiles we have measured the colours in wedges, at 15° from the major axis, with a width of 10°.

For this procedure to work the disk itself should contain relatively little extinction, since otherwise both sides are severely affected. A way to measure the extinction in the disk is to measure its radial colour gradient between a band that is affected by extinction and one that is not, e.g. $B$ and $K$ (Peletier *et al.* 1995). If the density ratio of dust and stars is more or less constant the central extinction will be much larger than the extinction in the outer parts, purely because of the difference in stellar density. In dusty galaxies this will cause the scale lengths in $B$ to be much larger than those in $K$. In Fig. 3 we plot scale length ratios between $B$ and $K$ as a function of galaxy type. The open squares are points from Peletier *et al.* (1994). In that paper also details are given about how the radial scale lengths have been determined. Fig. 3 is very instructive. One can see that the scale length ratios for galaxy types up to 2 (Sab) are almost never larger than 1.4. Scale length ratios for galaxies of type 6 and larger also are small. Only galaxies of type 3-5 are sometimes very dusty. This is in good agreement with the surface-brightness vs. inclination test (Valentijn 1990). An independent estimate for the scale length ratios due to stellar population gradients is 1.1 − 1.2 (Peletier *et al.* 1995). This means that disks of galaxies of type 2 or smaller are likely not to be very dusty, and so the effects of dust in this paper will in general not be very large. We may be reasonably confident to use the colours for stellar population measurements.

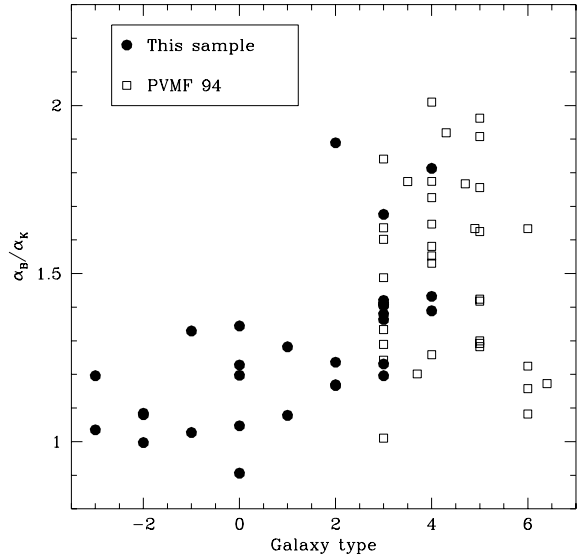

*Figure 3.*   Scale length ratios between B and K of galaxy disks as a function of galaxy type.

## 4. Stellar populations from dustfree colours

In Paper I a discussion is given of optical ($U - R$, $B - R$ and $R - I$) colours and colour gradients as a function of other parameters like bulge luminosity. It was found that about half the bulges have the same colour as elliptical galaxies of the same luminosity, and that the other half is bluer in $U - R$ and $B - R$. This was interpreted as a sign of the presence of younger stars, under the assumption that all ellipticals are old (15 − 17 Gyr). Now, with the combination of optical and near-infrared colours, we don't have to make that assumption any more and can directly determine the average age of the stars.

First we present, in Fig. 4, some colour-colour diagrams for the bulges. 30 galaxies are plotted, at $r_{\rm eff}/2$, or 5", if $r_{\rm eff} < 10$", but only 20 in the diagram in which $J$ is included, since 10 galaxies were not observed in $J$. Also included are lines of single burst models of constant age, with metallicity varying. These models are from Vazdekis *et al.* (1995) described elsewhere in these proceedings. The models are made by converting theoretical parameters like effective temperature and gravity to colours, choosing an initial mass function and integrating along a theoretical HR diagram, similar to Peletier (1989) and Worthey (1994). In this case however, we have taken a lot of care with the temperature–metallicity–colour calibration that the integrated models fit the data for giant elliptical galaxies. For example, for a given $B - V$ the $V - K$ given by Worthey (1994) is much too red to fit the giant elliptical galaxies. In Fig. 4 models are plotted for ages of 17, 12, 8, 4 and 1 Gyr. Along the lines metallicity ranges from Z=0.05 to Z=0.0004. One can see that only the diagrams that contain optical and optical-infrared colours can separate age and metallicity. $R - K$ and $J - K$ are both sensitive to more or less the same kind of stars, so in this diagram metallicity and age cannot be separated. The same holds for the $U - R$ vs. $B - R$ diagram. The systematic error in the model colours, due to uncertainties in the stellar evolution theory, or lack of template stars, seems to be on the order of 0.1 or 0.2 mag, although this value can be smaller, since the reddest bulges not necessarily have to be 17 Gyrs old. Despite these errors some galaxies in the $U - R$ vs. $R - K$ diagram cannot be fitted by a model of 17 Gyr, and for some the best fitting model only has an age of 1 Gyr. This does not mean that there is no underlying old stellar population present, but just that a much younger stellar population is dominant.

We see that most of the reddest, and also largest (Paper I) bulges are old, and that some have to be younger. It is also noteworthy that all bulges here are redder than $R - K = 2.4$. This means that the metallicity of all these bulges is larger than about 0.3 $Z_\odot$. This is probably due to the way the sample was chosen - bright nearby galaxies, automatically excluding all dwarfs, and galaxies with low metallicities.

The colour - colour diagrams for disks are very similar. The differences in colour between the bulge at 0.5 $r_{eff}$ and the disk at two scale lengths is much smaller than the range in colour indicated in Fig. 4. In Peletier & Balcells (1995) it is explained that this means that disks are at most 2-4 Gyr younger than bulges. The colours don't exclude an equal age, which would imply that the colour differences are due to metalicity gradients.

In Fig. 5 a plot is given of $R - K$ colour gradients in the bulges as a function of total $K$-band magnitude. Since $R - K$ is an indicator of old stellar populations measuring the bulge gradients in this colour is a good way to measure the metallicity gradients in bulges. Two different symbols are used for our galaxies. The filled circles indicate the galaxies of the dustfree sample of Paper I. The filed squares are those that may contain some dust. There are no significant differences to be seen between the $R - K$ gradients of both groups, showing that dust extinction is not a major factor influencing the gradients. The gradients are somewhat larger than the visual-infrared gradients of elliptical galaxies (Peletier et al. 1990, Silva & Elston 1994). Our average $R - K$ gradient is $\Delta(R - K)/\Delta(\log r) = -0.232$, and for $U - R$ $-0.399$. For ellipticals average values are $-0.14$ for $R - K$ and $-0.20$ for $U - R$, although not much data are available for $R - -K$. In Paper I we concluded from the similarity between gradients in the optical between bulges and disks that this is a strong argument to show that bulges and ellipticals formed in the same way, and that the subsequent disk-formation did not substantially affect the bulge. We can say the same now using the $R - K$ colour. A recent review by Minniti (1995) shows that possibly also the bulge of our own galaxy has a similarly small colour/metallicity gradient, even though earlier measurements by Terndrup (1988) were indicating large values. Fisher et al. (1995) find that for a sample of S0 galaxies the vertical gradients in the bulge in $B - R$ and in absorption line strength are much larger than the radial gradients. Although we don't have many galaxies in our sample that are really edge-on we also see hints for the same effect, and this might be the reason of Terndrup (1988)'s large metallicity gradient in our Bulge. Using colour - metallicity conversions by Vazdekis et al. (1995) we have converted our average $R - K$ and $U - R$ gradients to metallicity gradients, assuming constant age. From $R - K$ we find that $\Delta(\log Z) = -0.41$ per dex in radius, and from $U - R$ we find a value of $-0.42$.

Also plotted in Fig. 5 is the sample of Terndrup et al. (1994) of similar kinds of galaxies. Even though almost all $R - K$ gradients of our sample are negative, those of Terndrup et al. can be both positive and negative. We presume that the errors in that sample are larger than indicated by the errorbars, something which is possible, since they used a small detector covering a small field.

As a final point we would like to talk a bit about colours of disks as a function of galaxy type. Here the colours are the ones that have been determined in such a way as to minimize the effects of extinction (see above). 4 different colours are plotted in Fig. 6. The $J - K$ and $R - K$ colours here are indicators of the old underlying stellar population that constitutes most of the mass. $B - R$ and $U - R$ also depend on additional young stars. Noteworthy is the fact that there is not much change in the colour, when varying galaxy type. One can say that ellipticals, S0's and spirals up to type 2 (Sab's) have more or less the same colours, indicating few young stars and similar metallicities. Some galaxies of type 3 obviously suffer from extinction effects, while in all colours, but especially in $U - R$ galaxies of type 4 (Sbc's) are bluer than the others. This figure shows again that a considerable fraction of the stars in Sbc's and probably also later type spirals is young.

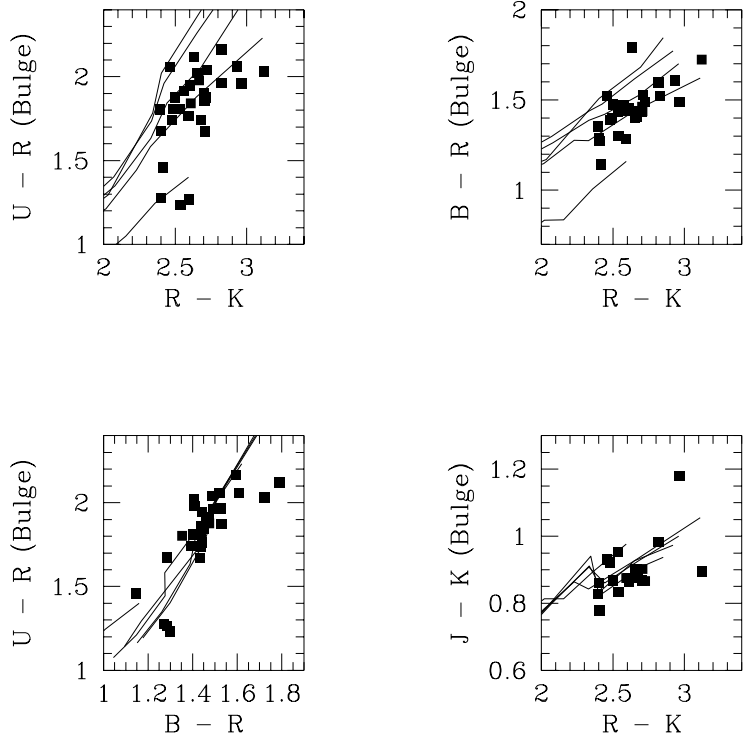

*Figure 4.* Colour-colour relations for the bulges. Also drawn are lines of constant age by Vazdekis *et al.* (1995). Plotted are ages 17, 12, 8, 4 and 1 Gyr.

## 5. Conclusions

The most important results that have been discussed in this paper are:

- By investigating colour gradients in disks between $B$ and $K$ for a sample of 70 galaxies we find that galaxies of type 3-5 are much more affected by extinction than are others. In the galaxies of other types, it is possible to find regions that are only slightly affected by extinction, and for which the colours can be used to study stellar populations.
- Using colour-colour diagrams of optical and optical-infrared colours, and comparing them with single age, single metallicity stellar population models, we find that a considerable fraction of the bulges in the sample has to be younger than elliptical galaxies, or contain a large fraction of stars that are younger than those in elliptical galaxies. The differences in colour from one galaxy to another can be much larger than the colour difference between the bulge and disk of the same galaxy. For that reason the inner disks of these galaxies cannot be more than 2-4 Gyrs younger than their bulges.
— Radial $R - K$ colour gradients of bulges are negative and small. Bulges generally are redder in the center than in their outer parts, and the average colour gradient is about 0.2 mag per dex. These values are due to metallicity, or possibly age gradients, but not due to extinction by dust. Metallicity gradients inferred from $R - K$ gradients are very similar to those calculated from $U - R$, indicating that the contribution from very young stars in these galaxies is small.
— Dustfree colours in disks are remarkably constant. Only for galaxies of type 3 and later disks are significantly bluer than those of S0's.

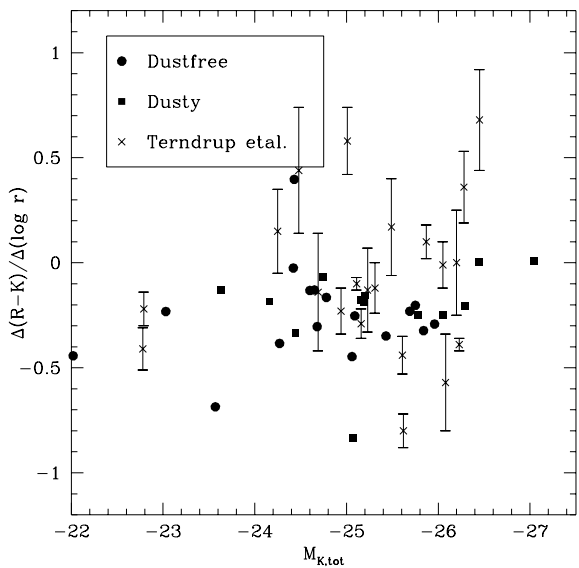

*Figure 5.*  $R - K$ gradients for the bulges. Also plotted are the data of Terndrup *et al.* (1994)

## Acknowledgments

We acknowledge the help Massimo Stiavelli for help with observations and Alejandro Vazdekis making available his stellar population models before publication.

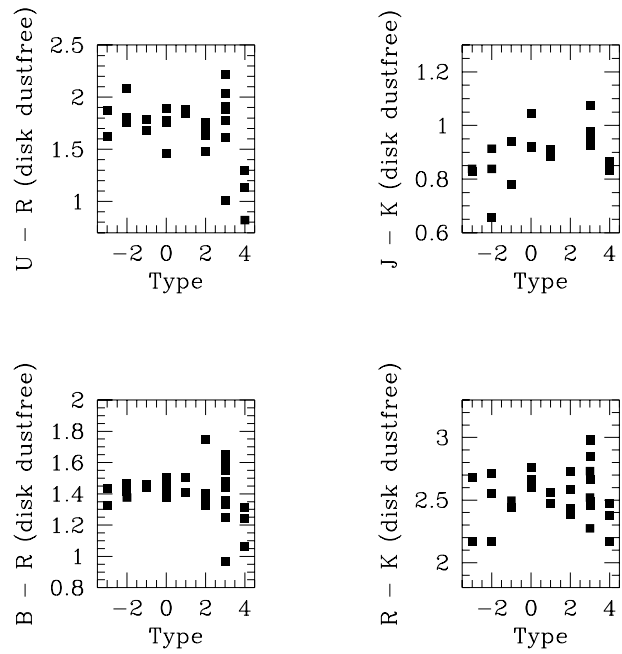

*Figure 6.* Disk colours as a function of galaxy type.